\documentclass[twocolumn,showpacs,pre,amsmath,amssymb,floatfix,superscriptaddress]{revtex4}
\usepackage{graphicx}
\usepackage{dcolumn}
\usepackage{bm}
\usepackage{hyperref} 
\usepackage{latexsym}
\usepackage{float}
\begin{document}

\title{Statistics of Weighted Tree-Like Networks}
\author{E.~Almaas}\email{ealmaas@nd.edu}  
\affiliation{Center for Network Research and Department of Physics, 
University of Notre Dame, Notre Dame, IN 46617, USA}
\author{P.~L.~Krapivsky}\email{paulk@bu.edu}
\author{S.~Redner}\email{redner@bu.edu}
\affiliation{Center for BioDynamics, Center for Polymer Studies, 
and Department of Physics, Boston University, Boston, MA 02215, USA}

\begin{abstract}
  
  We study the statistics of growing networks with a tree topology in which
  each link carries a weight $(k_i k_j)^\theta$, where $k_i$ and $k_j$ are
  the node degrees at the endpoints of link $ij$.  Network growth is governed
  by preferential attachment in which a newly-added node attaches to a node
  of degree $k$ with rate $A_k=k+\lambda$.  For general values of $\theta$
  and $\lambda$, we compute the total weight of a network as a function of
  the number of nodes $N$ and the distribution of link weights.  Generically,
  the total weight grows as $N$ for $\lambda>\theta-1$, and super-linearly
  otherwise.  The link weight distribution is predicted to have a power law
  form that is modified by a logarithmic correction for the case $\lambda=0$.
  We also determine the node strength, defined as the sum of the weights of
  the links that attach to the node, as function of $k$.  Using known results
  for degree correlations, we deduce the scaling of the node strength on $k$
  and $N$.

\end{abstract}
\pacs{02.50.Cw, 05.40.-a, 89.75.Hc}

\maketitle

\section{Introduction}

The recent interest in networks stems from the discovery that the node degree
distribution can have a power law form.  Here the node degree is defined as
the number of links that are connected to this node.  It is now well
documented that a wide variety of natural and man-made networks have
power-law degree distributions \cite{net-rev}.

In most previous studies, it has been implicitly assumed that the quality of
each link is identical.  However, there are many examples of networks in
which the weight of each link can be distinct.  For example, in scientific
collaborations, co-authors can have a variable number of joint publications.
Thus in the associated co-authorship network, a link between two individuals
represents the collaboration strength, such as the number of jointly-authored
publications \cite{New,B}.

Transportation networks represent another situation where the weight of each
link is variable.  In the specific example of airline route networks, links
between two airports represent the passenger capacity on this route
\cite{air}.  It has been found that the weight of the link (defined as the
number of available seats on flights between the two airports connected by
the link) is well-approximated by the products of the endpoint degrees raised
to a power \cite{orsay}.  A related line of research focuses on general
properties of flows in heterogeneous networks \cite{F1,F2}.

A third example where link weights are connected to network topology occurs
in the metabolic network of the bacterium {\it E.  coli}.  Here a link
between two metabolic substrates represents an enzymatic reaction, and the
weight of a link (defined as the metabolic flux rate) displays the same
functional dependence on the endpoint degrees as that of the world airline
network \cite{macdonald}.

There has also been much very recent work on various aspects of weighted
networks including characterizing paths on these networks \cite{NR},
structural properties \cite{hui,New2}, and the application to weighted
networks to diverse physical systems, such as earthquakes \cite{maya} and to
synchronization phenomena \cite{jurgen}.

In this work, we characterize basic properties of growing networks in which
the links possess variable weights \cite{BBV}.  We consider the generic
situation in which network growth is governed only by the node degrees, so
that the link weights are passive subsidiary variables.  (The comlementary
situation in which the link weights determine network growth has also been
analyzed, see, {\it e.g.}, Refs.~\cite{macdonald,NR,hui,BBV,AK}).  We define
the weight $w_{ij}$ of a link between nodes $i$ and $j$ in the network to be
the product of endpoint node degrees; that is, $w_{ij}=k_ik_j$.  More
generally, we study the general case where the link weight has the following
dependence on the endpoint degrees:
\begin{equation}
\label{theta}
w_{ij}=(k_ik_j)^\theta. 
\end{equation}
We show that this weighted network has a surprising dependence of the total
weight on the system size and an unusual scaling for the distribution of link
weights.  Many of the new results that we obtain follow naturally from
previous work on the degree correlations in growing networks \cite{KR}.

For the network growth mechanism, we employ preferential attachment, in which
a new node links to previous nodes with an attachment rate $A_k$ that depends
only on the degree of the ``target'' node \cite{BA}.  For simplicity, we
consider the situation where the newly-introduced node connects to a single
target.  Because of this topological restriction, the resulting network is a
tree, a feature that turns out to simplify our analysis.  To complete the
description of the system, we need to specify the attachment rate $A_k$.  We
will consider the Simon model \cite{simon}, where $A_k=k+\lambda$, with
$\lambda>-1$ but otherwise arbitrary.  For this growth mechanism the degree
distribution has a power law tail $\propto k^{-3-\lambda}$.

In the next section we compute the total network weight $W(N)$, defined as
the sum of the link weights $w_{ij}$, as a function of the total number of
nodes $N$.  The distribution of link weights is investigated in Sec.~III, and
the node strength, defined as the sum of the weights of all the links that
attach to the node, is studied in Sec.~IV.  We first focus on the case of the
strictly linear attachment rate where $A_k=k$ and then outline corresponding
results for the more general case of $A_k=k+\lambda$ with $\lambda>-1$ but
otherwise arbitrary.  Numerical simulation results and comparison to our
analytic predictions are given in Sec.~V.

\section{Total Network Weight} 

We denote by $N_{k,l}(N)$ the number of nodes of degree $k$ that attach to an
ancestor node of degree $l$.  This degree correlation function is
well-defined only for growth processes in which each node attaches to exactly
one previous node, so that the ancestor of each node is unique.  Notice that
the correlation function $N_{k,l}$ is not symmetric with respect to the end
nodes.  For graphs that are built sequentially by preferential attachment, we
are interested in the total weight of the network, defined by
\begin{equation}
\label{W}
W(N)=\sum_{k,l}kl\,N_{k,l}(N)\,. 
\end{equation}
That is, we first consider the case of $\theta = 1$ in Eq.~(\ref{theta}).

For growth processes in which the attachment rate $A_k$ does not grow faster
than linearly with $k$, the correlation function $N_{k,l}(N)$ grows linearly
with the total number of nodes $N$ (see \cite{KR}, and also \cite{rev} for a
brief review).  That is, $N_{k,l}(N)=Nn_{k,l}$ for large $N$. The explicit
form of the distribution $n_{kl}$ for the simplest case of the
Barab\'asi-Albert (BA) model ($\lambda=0$) is \cite{KR}
\begin{eqnarray}
\label{nkl-sol}
n_{k,l}&=&\frac{4(l-1)}{k(k+1)(k+l)(k+l+1)(k+l+2)}\nonumber\\
&+&\frac{12(l-1)}{k(k+l-1)(k+l)(k+l+1)(k+l+2)}\,.
\end{eqnarray}

Since $W(N)=N\sum kl\,n_{k,l}$, one might naively anticipate that the total
weight will scale linearly with $N$.  This expectation is erroneous because
the infinite sum $\sum kl\,n_{k,l}$ formally diverges, leading to an
additional $N$ dependence in $W(N)$.

To compute $\sum kl\,n_{k,l}$ we write $m=k+l$ and recast the sum as
\begin{eqnarray*}
\sum_{k,l}kl\,n_{k,l}&=&\sum_{m\geq 2} \frac{4}{m(m+1)(m+2)}
\sum_{l=2}^{m-1}\frac{l(l-1)}{m+1-l}\\
&+&\sum_{m\geq 2} \frac{12}{(m-1)m(m+1)(m+2)}
\sum_{l=2}^{m-1} l(l-1).
\end{eqnarray*}
For large $m$, the internal sums behave as
\begin{eqnarray*}
\sum_{l=2}^{m-1}\frac{l(l-1)}{m+1-l}&=&
m^2\ln m+{\cal O}(m^2), \\
\sum_{l=2}^{m-1} l(l-1)&=&\frac{1}{3}\,m^3+{\cal O}(m^2).
\end{eqnarray*}
Thus the asymptotically dominant part of the sum is
\begin{equation}
\label{sum}
\sum_{k,l}kl\,n_{k,l}\sim\sum_{m\geq 2} \frac{4\,\ln m}{m}
+\sum_{m\geq 2} {\cal O}(m^{-1})
\end{equation}

To evaluate these sums explicitly, we need to impose a cutoff in the upper
limit.  For the case $\lambda=0$, we therefore use the fact that the maximal
degree $k_{\rm max}$ scales as $N^{1/2}$ \cite{BA,KR-finite}.  This result is
easy to derive by utilizing $N_k\sim 4Nk^{-3}$ and estimating the
maximal degree from the extremal criterion that in a network of $N$ nodes,
there will be a single node whose degree in the range $(k_{\rm max},\infty)$.
This criterion corresponds to the condition $\sum_{k\geq k_{\rm max}}N_k=1$.
Thus we cut off the sums in (\ref{sum}) at $m_{\rm max}\sim (2N)^{1/2}$ to give
\begin{equation}
\label{sum-N}
\sum_{k,l}kl\,n_{k,l}=\frac{1}{2}\,(\ln N)^2 +{\cal O}(\ln N).
\end{equation}
Since the leading sum in (\ref{sum}) diverges logarithmically, the mere
knowledge of the scale of the cutoff is sufficient to determine the exact
asymptotics.  With this result for the internal sums, we thereby find, from
(\ref{W}),
\begin{equation}
\label{WBA}
W(N)=\frac{1}{2}\,\,N(\ln N)^2 +{\cal O}(N\ln N).
\end{equation}

A similar computation for the Simon model (where $A_k=k+\lambda$) with arbitrary
$\lambda$ is difficult to perform exactly because the expression for
$n_{k,l}$ is an infinite series \cite{KR} rather than a rational function as
in Eq.~(\ref{nkl-sol}).  However, the dominant contribution to the sum
$\sum_{k,l}kl\,n_{k,l}$ is concentrated in the region $1\ll k\ll l$, where
$n_{k,l}$ admits the simple asymptotic form \cite{KR}
\begin{equation}
\label{nkl-gen}
n_{k,l}\sim k^{-2}l^{-2-\lambda}\,. 
\end{equation}
Therefore
\begin{eqnarray}
\label{klnkl}
\sum_{k,l}kl\,n_{k,l}\sim \sum l^{-1-\lambda}\sum_{k\ll l} k^{-1}
\sim \sum l^{-1-\lambda}\,\ln l.
\end{eqnarray}
For $\lambda>0$ the sum converges, while for $\lambda\leq 0$ the sum formally
diverges and we must again determine an upper cutoff.  Applying the
previously-mentioned extremal argument for the maximal node degree for
general values of $\lambda$, we now find $k_{\rm max}\propto N^{1/(2+\lambda)}$
\cite{KR-finite}.  Thus we conclude that the last sum in Eq.~(\ref{klnkl})
scales as $N^{-\lambda/(2+\lambda)}\,\ln N$ for $-1<\lambda<0$.  In summary,
\begin{equation}
\label{WS}
W(N)\propto 
\begin{cases}
N                        & \lambda>0;\cr
N\,(\ln N)^2             & \lambda=0;\cr
N^{2/(2+\lambda)}\ln N   & \lambda<0.
\end{cases} 
\end{equation}

Consider now the situation where link weight depends on its endpoint degrees
according to Eq.~(\ref{theta}).  We treat only the case $\theta<1$ because
this regime corresponds to real networks \cite{orsay,macdonald}.  Using the
asymptotic behavior for $n_{k,l}$ given in Eq.~(\ref{nkl-gen}), we then have
\begin{eqnarray*}
\sum_{k,l}(kl)^\theta \,n_{k,l}\sim \sum l^{-2-\lambda+\theta},
\end{eqnarray*}
which converges for $\lambda>\theta-1$ and diverges otherwise.  Thus in the
range $\theta$ strictly less than 1, the total weight of the network scales
according to
\begin{equation}
\label{W-theta}
W(N)\sim 
\begin{cases}
N                                  & \lambda>\theta-1;\cr
N \ln N                            & \lambda=\theta-1;\cr
N^{(1+\theta)/(2+\lambda)}         & \lambda<\theta-1.
\end{cases} 
\end{equation}

\section{Link Weight Distribution} 

In this section, we consider exclusively the case of $\theta=1$.  We define
the link weight distribution $\Pi(w)$ as the fraction of links whose weight
equals $w$.  That is,
\begin{equation}
\label{Pi-def}
\Pi(w)=\sum_{\stackrel{k,l}{kl=w}}\,n_{k,l}
\end{equation}
This natural definition leads to erratic behavior in the distribution, as the
arithmetic nature of $w$ plays a significant role.  If $w$ is prime, then the
only possible degrees of the end nodes are $1$ and $w$, while if $w$ has many
divisors, there are more possibilities for the degrees of the end nodes.
Thus in addition to an expected global smooth dependence of $\Pi(w)$ on $w$,
there will be superimposed fluctuations related to the divisibility of $w$.
These fluctuations will disappear if we average $\Pi(w)$ over a range much
less than $w$ and much larger than the average distance between successive
primes (which grows as $\ln w$ \cite{Hardy}).  This suggests that we also
consider a smoothed distribution ${\mathcal P}(w)$ defined by
\begin{equation}
\label{Pi-average}
\mathcal{P}(w)=w^{-a}\sum_{w\leq kl\leq w+w^a}\,n_{k,l},
\end{equation}
with $a$ in the range $0<a<1$.  For sufficiently large $w$ this smoothed
distribution should be monotonic and independent of $a$.

To determine the behavior of the link weight distribution for $w\gg 1$, we
again start with the case of the BA model where $\lambda=0$.  Using
(\ref{nkl-sol}) and writing $l=w/k$ we obtain, for large $w$,
\begin{equation}
\label{sum-div}
\Pi(w)
\to \sum_{k|w}\left[\frac{4w}{(k^2+w)^3}+\frac{12w k^2}{(k^2+w)^4}\right].
\end{equation}
The notation $k|w$ means that this sum runs over all values of $k$ that are
divisors of $w$.  For large $w$, the asymptotic behavior of the summand
scales as $w^{-2}$ times the average number of divisors of $w$.  Using the
celebrated Dirichlet formula (see {\it e.g.}, \cite{Hardy}), the total number
of divisors of all integers from 1 to $w$ is
\begin{equation}
\label{Dirichlet}
\sum_{x=1}^w d(x)=w\ln w +(2\gamma-1)w+\ldots,\nonumber
\end{equation}
where $d(x)$ is the number of divisors of the integer $x$ and $\gamma\cong
0.577215$ is the Euler-Mascheroni constant.  Thus the average number of
divisors of $w$ grows logarithmically with $w$; more precisely, by
differentiating the above formula with respect to $w$, we have $\langle
d(w)\rangle\to\ln w + 2\gamma$.  Thus the tail of the weight distribution is
\begin{equation}
\label{Pi-large}
{\mathcal{P}}(w)\sim w^{-2}\,\ln w\qquad {\rm as}\quad w\gg 1.
\end{equation}

Consider now the general case of arbitrary $\lambda$ (but with the
necessary constraint $\lambda>-1$ to ensure a finite average degree).
In this case, we have to combine two separate contributions to
determine the link weight distribution.  First, using $kl=w$, we
re-write Eq.~(\ref{nkl-gen}) as
\begin{equation}
\label{nkl-}
n_{k,l}\sim k^{\lambda}\,w^{-2-\lambda}. 
\end{equation}
This asymptotic behavior holds for $k<l$, so it can be used to
estimate the part of the sum in Eq.~(\ref{Pi-def}) that runs over the
divisors $k|w$ such that $k<\sqrt{w}$.  The corresponding contribution
to the weight distribution is therefore
\begin{equation}
\label{-}
{\mathcal P}_<(w)\sim w^{-2-\lambda} \sigma_\lambda(\sqrt{w}),
\end{equation}
where $\sigma_\lambda(n)$ is the generalized divisor function of $n$ defined
as
\begin{equation}
\label{divisor}
\sigma_\lambda(n)=\sum_{f|n}f^\lambda\,,
\end{equation}
where the sum runs over all divisors $f$ of the integer $n$.  This generalized
divisor function exhibits, on average, a power-law growth in $n$ for
$\lambda>0$, $\sigma_\lambda(n)\sim n^\lambda$, and approaches a constant for
$\lambda<0$ \cite{Sch}.  As a result, the asymptotic behavior of ${\mathcal
  P}_<(w)$ undergoes a qualitative change when $\lambda$ passes through zero.
Using the result for the generalized divisor function in Eq.~(\ref{-}), we
find, for the contribution to the link weight distribution from divisors of
$w$ that are smaller than $\sqrt{w}$:
\begin{equation}
\label{P-summary}
{\mathcal P}_<(w)\sim 
\begin{cases}
w^{-2-\lambda/2} & \lambda>0, \cr
w^{-2-\lambda} & \lambda<0.
\end{cases}
\end{equation}

A second contribution to the weight distribution arises from the range $k>l$.
(For the following arguments we use the smoothed weight distribution of
Eq.~(\ref{Pi-average}).  Correspondingly, the generalized divisor function
$\sigma_\lambda(n)$ should also be averaged in the same way as that for the
weight distribution in Eq.~(\ref{Pi-average}).)~ In this case, the asymptotic
behavior of the distribution $n_{k,l}$ is \cite{KR}
\begin{equation}
\label{nkl+}
n_{k,l}\sim l\,^{6+3\lambda}\,w^{-5-2\lambda} 
\end{equation}
We use this asymptotics to estimate the part of the sum in Eq.~(\ref{Pi-def})
that runs over the divisors $k|w$ such that $k>\sqrt{w}$, or equivalently
over divisors $l|w$ such that $l<\sqrt{w}$.  The corresponding contribution
to the weight distribution is
\begin{equation}
\label{+}
{\mathcal P}_>(w)\sim w^{-5-2\lambda} \sigma_{6+3\lambda}(\sqrt{w})
\end{equation}
Since $6+3\lambda>0$ (in fact, the stronger inequality $6+3\lambda>3$ holds),
the generalized divisor function behaves as
$\sigma_{6+3\lambda}(\sqrt{w})\sim w^{3+3\lambda/2}$.  Thus
\begin{equation}
\label{++}
{\mathcal P}_>(w)\sim w^{-2-\lambda/2}.
\end{equation}

We therefore find that the large-$w$ tail of the complete weight distribution
${\mathcal P}(w)={\mathcal P}_<(w)+{\mathcal P}_>(w)$ is dominated by
${\mathcal P}_<(w)$.  Thus summarizing our results, the link weight
distribution is
\begin{equation}
\label{Weight}
{\mathcal{P}}(w)\sim 
\begin{cases}
w^{-2-\lambda/2}   & \lambda>0;\cr
w^{-2}\,\ln w      & \lambda=0;\cr
w^{-2-\lambda}     & \lambda<0.
\end{cases} 
\end{equation}
The average weight per link $\langle w\rangle=\int dw\,w\,\Pi(w)$ should be
equal $W(N)/N$.  In computing $\langle w\rangle$ using Eq.~(\ref{Weight}), we
find agreement with Eq.~(\ref{WS}) only for $\lambda\geq 0$.  The reason for
the disagreement in the range $-1<\lambda<0$ is unclear.  Part of difficulty
may lie in the fact that the link weight distribution exhibits an anomalous
feature that is not simply a power-law behavior (see Sec.~V).

\section{Node Strength}

Many real networks display a significant correlation between the degrees of
adjacent nodes.  There are examples where there is a bias for nodes to be
connected to other nodes with either similar (assortative) or different
(disassortative) degree \cite{newman}.  This tendency to select nodes with a
consonant property is often expressed as a Pearson correlation coefficient of
the network.  For networks where the weights are correlated with the
topology, we investigate the effects of correlations by studying the node
strength defined as \cite{orsay}
\begin{equation}
s_i ~=~ \sum_{\langle j i\rangle} w_{ij}, \label{eq:s}
\end{equation}
where the sum runs over all nodes $j$ that are linked to node $i$.  Let us
first consider the simplest link weight function where $w_{ij}$ is just the
product of the endpoint node degrees.  Then the strength $s(k)$ of a node of
degree $k$ is simply $k$ times the sum of the degrees of the neighboring
nodes.  Since there are $k$ neighbors, $s(k)$ is then $k^2$ times the average
degree of the neighboring nodes if we ignore correlations between the degrees
of neighboring nodes.  Now the average node degree equals twice the number of
links divided by the number of nodes; this simply equals 2 for a network with
a tree topology.  Hence the node strength $s(k)$ scales as
\begin{equation}
\label{simple}
s(k)\sim k^2\,.
\end{equation}

Because node degrees are actually correlated, we now investigate their
influence on the simple prediction of Eq.~(\ref{simple}).  To include the
effect of node correlations on the behavior of the node strength $s(k)$, we
start by splitting the contributions to $s(k)$ into two parts: one from
incoming links and another from the outgoing link.  This gives
\begin{equation}
\label{eq:sk}
s(k)=s_{\rm in}(k)+s_{\rm out}(k)
=(k-1)k\,\langle j_{k}\rangle_{\rm in} + k\,\langle l_{k}\rangle_{\rm out}\,.
\end{equation}
The first term accounts for the fact that there are $k-1$ incoming links,
each of which has weight $k\, j_k$, where $j_k$ denotes the degree of a
daughter node when the initial node has degree $k$.  Similarly, the second
term accounts for the single ancestor node with degree $l_k$.  We now
consider these two contributions to $s(k)$ separately.

For the contribution due to the ancestor node, we use the fact that its
average degree $\langle l_{k}\rangle_{\rm out}$ is given by
\begin{equation}
\label{eq:lk-out}
\langle l_{k}\rangle_{\rm out}=\frac{\sum_{l\geq 1} l\,n_{k,l}}{\sum_{l\geq 1} n_{k,l}}.
\end{equation}
The normalization factor is just the degree distribution 
\begin{equation*}
\sum_{l\geq 1} n_{k,l}=n_k=\frac{4}{k(k+1)(k+2)}.
\end{equation*}
Using this, together with Eq.~(\ref{nkl-sol}), we obtain (again using the
shorthand notation $m=k+l$)
\begin{eqnarray*}
&&\langle l_{k}\rangle_{\rm out}=(k+2)\sum_{l\geq 1}
\frac{l(l-1)}{m(m+1)(m+2)}\\
&&+3(k+1)(k+2)\sum_{l\geq 1}\frac{l(l-1)}{(m-1)m(m+1)(m+2)}\,.
\end{eqnarray*}
The first sum on the right-hand side is logarithmically divergent.  Using the
fact the upper cutoff is $\sqrt{N}$ and replacing the sum by an integral, we
find that for $k\gg 1$ the first sum is asymptotic to $\ln \sqrt{N}-\ln k$.
Similarly we find that the second sum is asymptotic to ${1}/({3k})$.
Therefore
\begin{eqnarray}
\label{eq:sout}
s_{\rm out}(k)= k\,\langle l_k\rangle_{\rm out}\to k^2  
\left(\ln \sqrt{N} -\ln k\right)\,.
\end{eqnarray}

For the contribution to $s(k)$ from the daughter nodes, we need the average
degree of these nodes.  This is given by
\begin{equation}
\label{eq:jk-in}
\langle j_{k}\rangle_{\rm in}=\frac{\sum_{j\geq 1} j\,n_{j,k}}{\sum_{j\geq 1} n_{j,k}}\,.
\end{equation}
The denominator is found from 
\begin{equation*}
\sum_{j\geq 1} n_{j,k}=(k-1)\,n_k=\frac{4\,(k-1)}{k(k+1)(k+2)}.
\end{equation*}
This equality is somewhat subtle.  A node of degree $k$ has $k-1$ incoming
daughter nodes.  If we sum the two-point correlations functions for all these
daughter nodes, we overcount the number of nodes of degree $k$ by exactly the
factor $k-1$.  The numerator can be computed straightforwardly but tediously
by following similar steps to those preceding Eq.~(\ref{eq:sout}) to evaluate
the sums.  The resulting asymptotic behavior has the simple form
$4\,k^{-2}\,\ln k$.  Thus we have
\begin{eqnarray}
\label{eq:sin}
s_{\rm in}(k)= (k-1)k\,\langle j_{k}\rangle_{\rm in}\to k^2 \ln k
\end{eqnarray}
Consequently, for $k \gg 1$ the node strength scales 
\begin{equation}
\label{eq:s-tot}
s(k) \to \frac{1}{2}\, k^2 \ln N\,. 
\end{equation}

A note of warning is in order.  The above computation of the contribution of
the outgoing link to the node strength is {\em exact} --- both $s_{\rm
out}(k) =k\,\langle l_k\rangle_{\rm out}$ and Eq.~(\ref{eq:lk-out}) are
manifestly correct, and therefore the knowledge of $n_{k,l}$ suffices to
determine $s_{\rm out}(k)$.  On the other hand, an exact calculation of the
contribution of incoming links requires knowing the many-body correlation
function $n(j_1,\ldots,j_{k-1})$.  This quantity gives the fraction of nodes
of degree $k$ whose daughter nodes have degrees $j_1,\ldots,j_{k-1}$.  This
exact contribution has the form
\begin{eqnarray*}
s_{\rm in}(k)&=&k\,\langle j_1+\ldots +j_{k-1}\rangle_{\rm in}\,,\\ 
&=&k\,
\frac{\sum_{j_1,\ldots,j_{k-1}}(j_1+\ldots +j_{k-1})\,n(j_1,\ldots,j_{k-1})}
{\sum_{j_1,\ldots,j_{k-1}}n(j_1,\ldots,j_{k-1})}\,.
\end{eqnarray*}
Unfortunately, we do not know $n(j_1,\ldots,j_{k-1})$, so we cannot compute
$s_{\rm in}(k)$ exactly.  Equations (\ref{eq:sk}) and (\ref{eq:jk-in}), which
involve only the known degree correlation $n_{k,l}$, are based on the
assumption that the degrees of the daughter nodes are uncorrelated.  Although
the correlation between the degrees of daughter nodes is certainly smaller
than the degree correlation between the daughter and mother nodes, it is not
clear that we can ignore these correlations.

Consider now the general linear attachment kernel $A_k = k + \lambda$ with
$\lambda > -1$ and choose the weight function $w_{ij}=(k_ik_j)^\theta$ with
$\theta <1$.  We now compute the contribution of the outgoing link to the
node strength.  We have
\begin{equation}
\label{sk-out}
s_{\rm out}(k)=\frac{k^\theta}{n_k}\,\sum_{l\geq 1} l^\theta\,n_{k,l}
\end{equation}
In performing this sum, we use the asymptotics $n_k\sim k^{-3-\lambda}$ for
$k\gg 1$ and the results of \cite{KR} for the asymptotics of $n_{k,l}$,
\begin{eqnarray*}
n_{k,l} &\sim& k^{-5-2\lambda}\, l^{1+\lambda}\qquad {\rm for}\quad 1\ll l\ll k\, \\
n_{k,l}&\sim& k^{-2}l^{-2-\lambda}\qquad \quad {\rm for}\quad 1\ll k\ll l\,.
\end{eqnarray*}
We estimate the formally diverging sums by using the cutoff
$N^{1/(2+\lambda)}$.  We thus find
\begin{eqnarray*}
s_{\rm out}(k)\sim 
\begin{cases}
 k^{2\theta}                                  & \lambda>\theta-1;\cr
 k^{2+2\lambda}(\ln N^{1/(2+\lambda)}-\ln k)  & \lambda=\theta-1;\cr
 k^{1+\lambda+\theta}\,N^{(\theta-1-\lambda)/(2+\lambda)} - k^{2\theta}   & \lambda<\theta-1.
\end{cases} 
\end{eqnarray*}
Finally, we specialize this prediction to the simplest weight function
$w_{ij}=k_ik_j$, as we report simulation results only in this particular case
in the next section:
\begin{eqnarray}
s_{\rm out}(k)\sim 
\begin{cases}
 k^{2}                                           & \lambda>0;\cr
 k^{2}(\ln \sqrt{N}-\ln k)                       & \lambda=0;\cr
 k^{2+\lambda}\,N^{-\lambda/(2+\lambda)} - k^2   & \lambda<0.
\end{cases} 
\label{eq:sgen}
\end{eqnarray}

\section{Simulation Results}
 
We now present numerical results for the total weight of a network and the
underlying weight and strength distributions.  To generate the network, we
use the redirection method, as discussed in Ref.~\cite{KR}.  In this
approach, we attach a new node to a randomly-selected target node with
probability $1-r$ and attach to the direct ancestor of this target with
probability $r$.  This method is both extremely simple, because the target
node is randomly selected, and efficient, since the time to build a network
of $N$ nodes scales linearly with $N$.
 
As discussed in Ref.~\cite{KR}, this redirection rule is equivalent to a
network in which the attachment rate $A_k$ to a node of degree $k$ equals
$k+\lambda$, with $\lambda=\frac{1}{r}-2$.  The attachment rate $k+\lambda$
then leads to a degree distribution $n_k\sim k^{-\nu}$, with $\nu=3+\lambda$
\cite{KR}.  Thus by implementing redirection, we can generate scale-free
networks with degree distribution exponent anywhere in the range
$(2,\infty)$.  As a parenthetical technical point, each node in the initial
network must have a unique ancestor to ensure that redirection weights each
network realization correctly \cite{KR-finite}.  In our simulations, we chose
the starting network to be a triangle in which each node points cyclically to
its nearest neighbor.  This detail about initial condition does not affect
the asymptotic form of the degree distribution.

\begin{figure}[tbp] 
  \vspace*{0.cm}
  \includegraphics*[width=8cm]{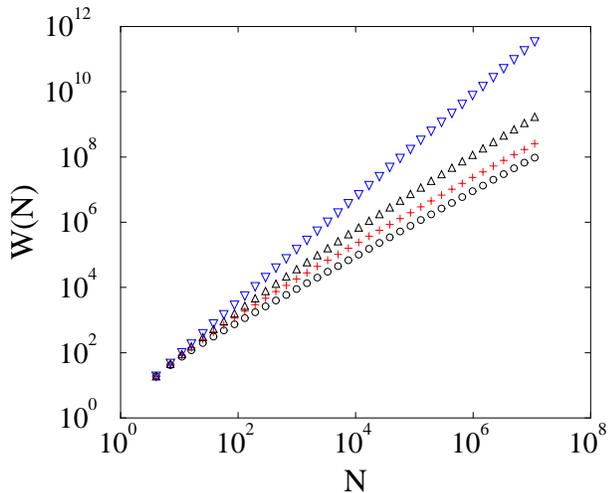}
 \caption{Average network weight $W(N)$ as a function of $N$ for the
   representative cases $\lambda =100, 1, 0$, and $-2/3$,
   $(\circ,+,\Delta,\nabla)$, respectively.}
 \label{WN}
\end{figure}

\begin{figure}[tbp] 
  \vspace*{0.cm}
  \includegraphics[width=7cm]{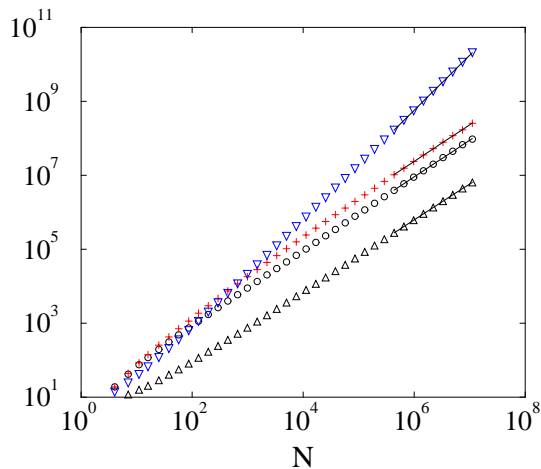}
 \caption{Same as in Fig.~\ref{WN}, except that the data for
   $\lambda=0$ and $\lambda<0$ has been divided by $(\ln N)^2$ and by $\ln N$,
   respectively.  The solid lines are best fits to the data for $N>4\times
   10^5$.  The slopes of these lines are 0.990, 0.988, 0.969, and $1.489$ for
   $\lambda=100, 1, 0$, and $-2/3$, respectively.}
 \label{W-div}
\end{figure}

Figure \ref{WN} shows the average network weight as a function of the total
number of nodes $N$.  These results are based on 100 realizations of the
network that each contain up to $N_{\rm max} = (1.5)^{40} \approx 1.1\times
10^7$ nodes.  On a double logarithmic scale, the data for $W(N)$ versus $N$
is quite linear for $\lambda>0$, but there is a very small systematic
curvature in the data for $\lambda\leq 0$.  As suggested by Eq.~(\ref{WS}),
we therefore divide the data for $W(N)$ by $(\ln N)^2$ for $\lambda=0$, or by
$\ln N$ for $\lambda<0$.
 
The resulting data shown in Fig.~\ref{W-div} are now visually more linear on
a double logarithmic scale.  Power-law fits to large-$N$ subranges of the
data give effective exponents that are only weakly dependent on the subrange.
For the case $\lambda=-2/3$, the effective exponent changes from $1.451$ to
$1.487$ as the lower limit of the fit range $N_{\rm min}$ is increased from
4 to $4\times 10^5$ while $N_{\rm max}$ remains fixed.  This behavior
suggests that the total network weight scales as $N^{3/2}\,\ln N$, as
predicted by Eq.~(\ref{WS}).  Similarly, for $\lambda=0$, the effective
exponent changes from 0.938 to 0.969 when the above fitting protocol is
applied.  Again, the data is consistent with $W(N)\sim N\,(\ln N)^2$ for
$\lambda=0$.  For $\lambda>0$, the variation in the effective exponent is
smaller still and it is evident that $W(N)$ varies linearly with $N$.

\begin{figure}[!ht] 
  \vspace*{0.cm}
  \includegraphics*[width=8cm]{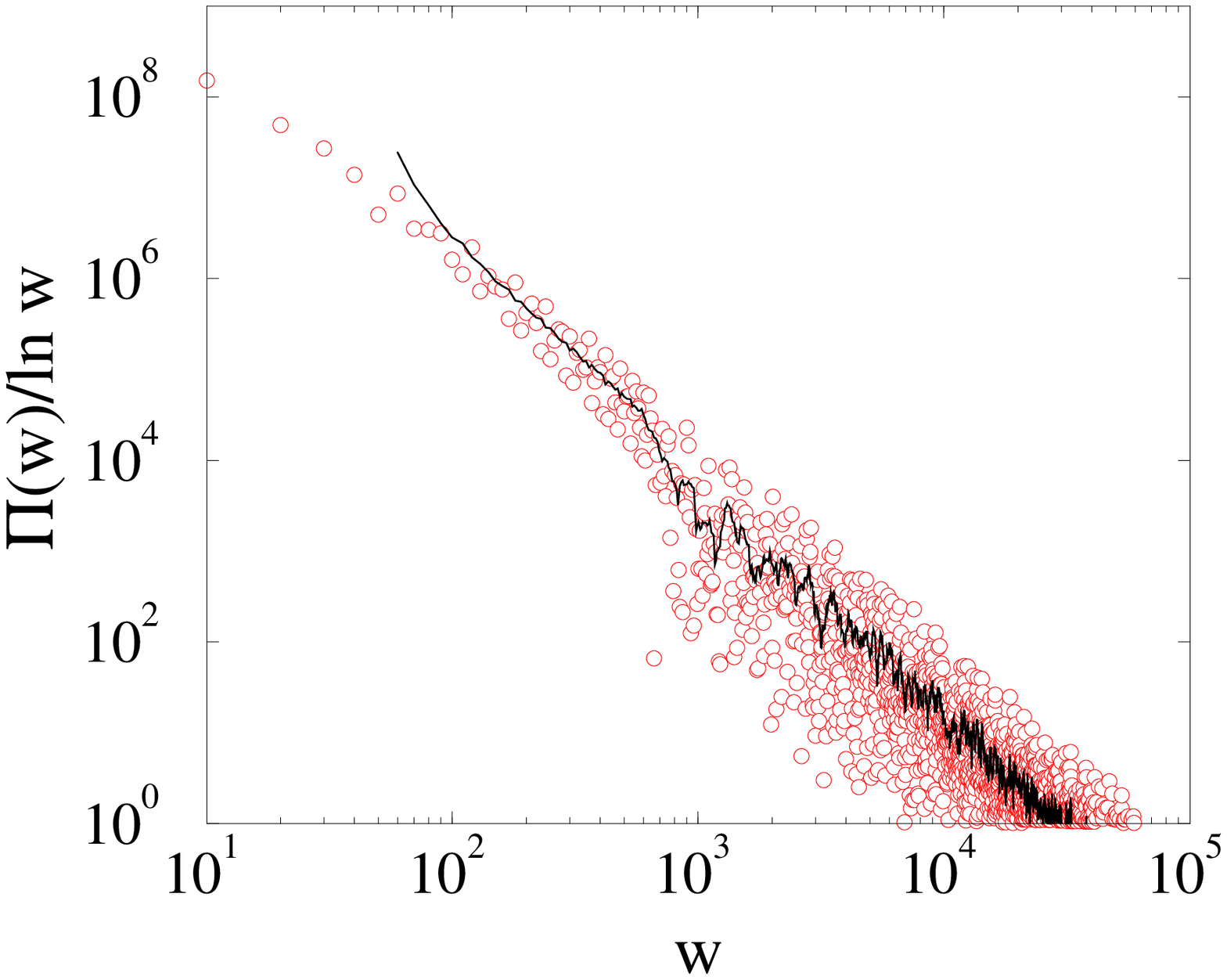}
  \includegraphics*[width=8cm]{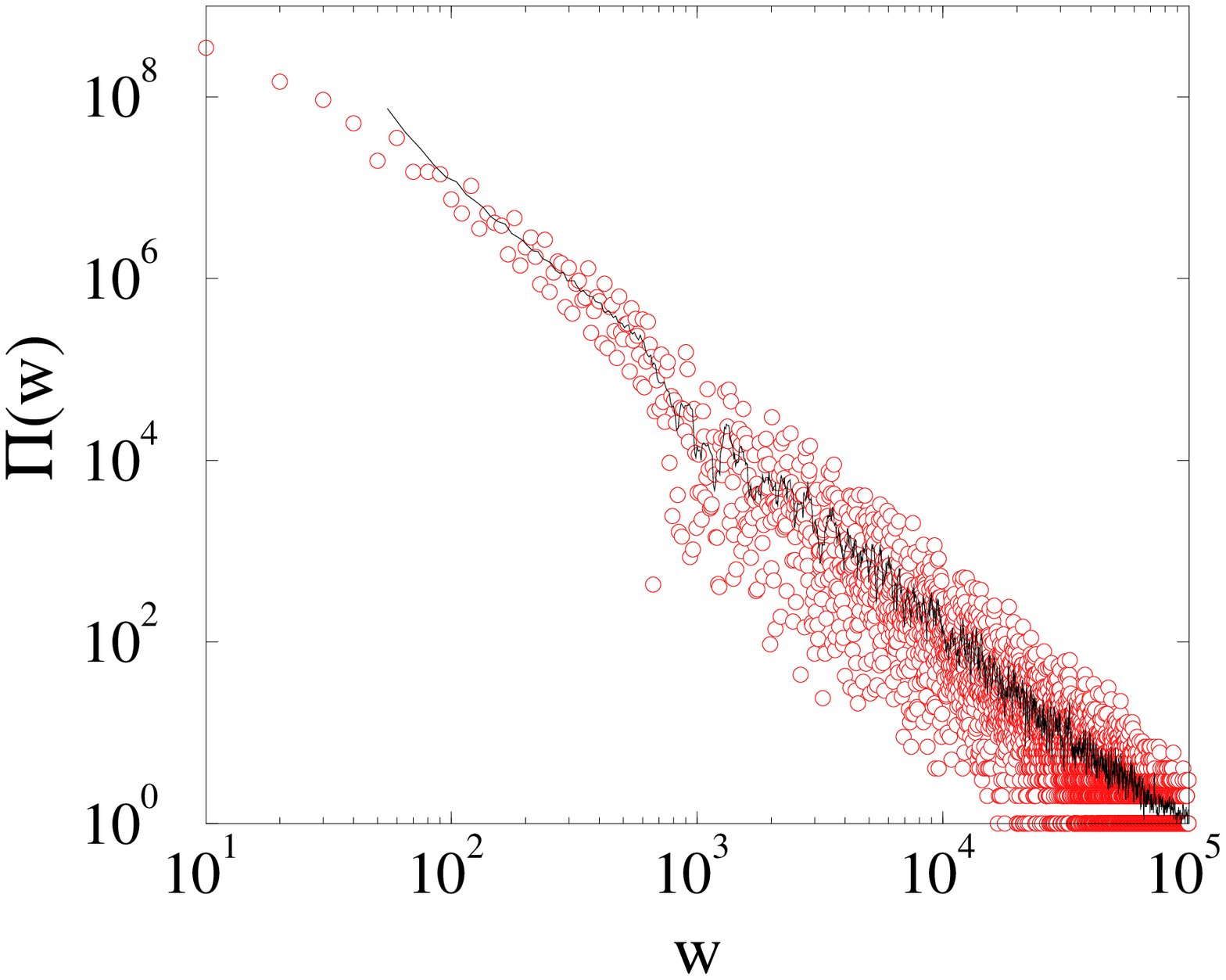}
  \includegraphics*[width=8cm]{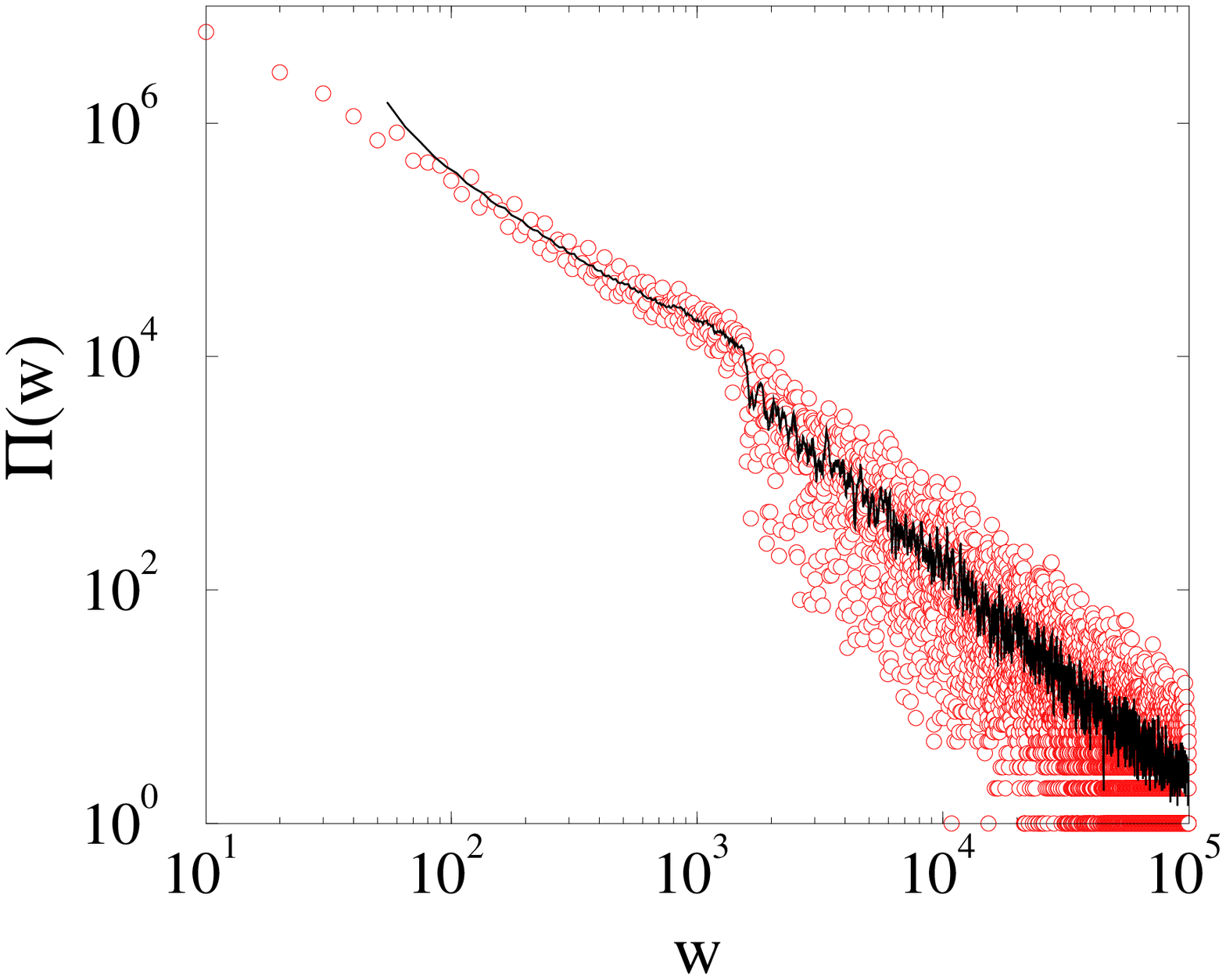}
 \caption{Distribution of link weights $\Pi(w)$ versus $w$ for $\lambda=0$
   (top), $\lambda=1$ (middle) and $\lambda=-0.5$ (bottom).  Every tenth
   point is plotted and the line is a 10-point average.  The quantity
   $\Pi(w)/\ln w$ appears in the top plot.}
 \label{lw}
\end{figure}

In Fig.~\ref{lw}, we show our data for the distribution of link weights for
the representative cases of $\lambda=0$, $1$, and $-0.5$.  The data shown are
based on $10^4$ network realizations with $N=10^6$ for $\lambda=0$ and 1,
while $N$ is limited to 25,000 for the $\lambda=-1/2$ due to the broadness of
the link weight distribution.  The data shows considerable small-scale
variation because of the previously-mentioned fluctuations in the number of
divisors of the integers; thus we also examined the locally averaged
distribution given by Eq.~(\ref{Pi-average}).  While this construction
smooths the distribution, exponent estimates based on the smoothed
distribution are nearly identical to those found for the raw distribution
($\Pi(w)$ in Eq.~(\ref{Pi-def})) and thus we quote results based on the
analysis of the latter.

For $\lambda=0$, a power-law fit to the data for $\Pi(w)/\ln w$ gives
$\Pi(w)/\ln w\sim w^{-\chi}$, with $\chi\approx 2.3$, compared to the
theoretical prediction from (\ref{Weight}) of $\chi=2$.  For $\lambda=1$, we
find $\Pi(w)\sim w^{-\chi}$, with $\chi\approx 2.23$, while the theoretical
prediction from (\ref{Weight}) is $\chi=2.5$.  Finally, for $\lambda=-0.5$, a
simple power law fit to the raw distribution in Fig.~\ref{lw} is clearly
an oversimplification.  However, a power-law fit to all the data gives
$\chi=1.39$, close to the theoretical prediction of $\chi=1.5$.  Overall, the
agreement between the predictions of Eq.~(\ref{Weight}) and simulation
results is surprisingly good, given the vagaries of these distributions.
 
\begin{figure}[tbp] 
 \vspace*{0.cm}
 \includegraphics*[width=8cm]{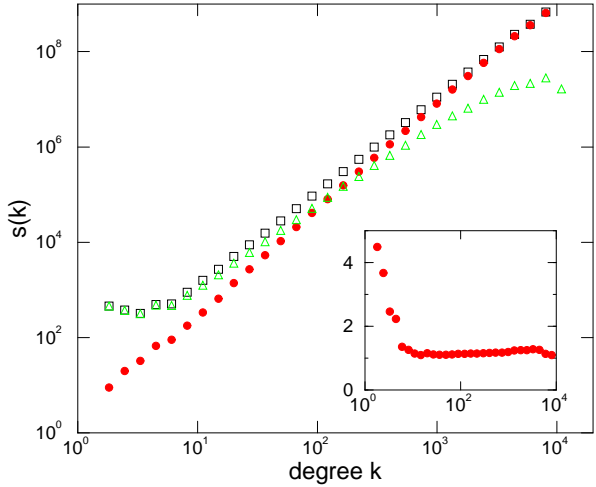}
  \caption{The node strength $s(k)$ versus $k$ for a network of $N=10^7$
    nodes with $\lambda=0$, and the contributions to this strength from
    incoming and outgoing links, as given by the two terms in
    Eq.~(\ref{eq:sk}) ($\square,\bullet,\Delta$ respectively).  Inset:
    $s(k)_{\rm in}$ divided by $k^2 \ln k$.}
  \label{strength1}
\end{figure}

\begin{figure}[tbp] 
  \vspace*{0.cm}
  \includegraphics*[width=8cm]{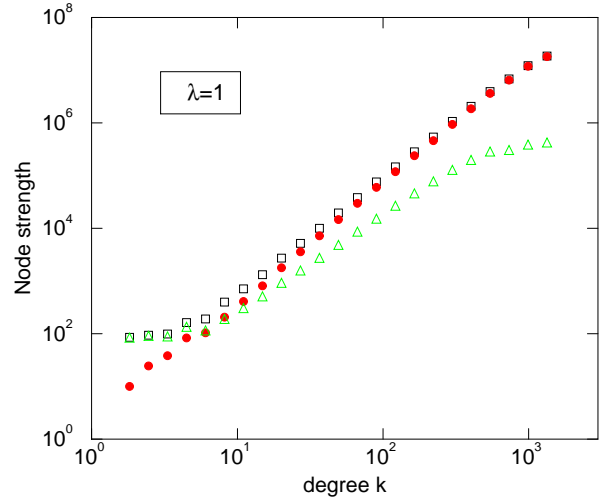}
  \includegraphics*[width=8.3cm]{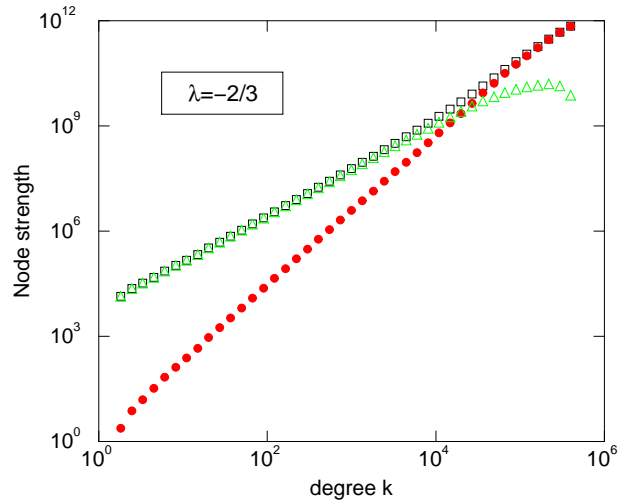}
  \caption{The node strength $s(k)$ and its contribution from incoming
    and outgoing links ($\square,\bullet,\Delta$ respectively) for the cases
    of $\lambda=1$ (upper) and $-2/3$ (lower), both with $N=10^7$.}
  \label{strength2}
\end{figure}

We also numerically calculate the node strength as function of the degree
$k$.  In Fig.~\ref{strength1} we show the results for $\lambda=0$ for a
network of $N=10^7$ nodes.  As a check of Eq.~(\ref{eq:sout}), the
contribution to the node strength from the outgoing links was fit with form
$s_{\rm out}(k) \sim k^\beta (\ln k_{\rm max} - \ln k)$.  Using a non-linear
curve fitting package, we obtain $\beta=1.9$ and $k_{\rm max}=13,000$.  This
value of $k_{\rm max}$ is very close to what we observed numerically (10,900)
for the maximum degree over all 1000 realizations of the network.  Overall,
the data for $s_{\rm out}(k)$ agrees reasonably well with the theoretical
prediction from Eq.~(\ref{eq:sout}) with $\beta = 2$.

In Fig.~\ref{strength2} we show the node strength for two other
representative values of $\lambda$.  For the case of $\lambda=1$, both
$s(k)$ and $s_{\rm out}(k)$ grow as $k^2$, as predicted by
Eqs.~(\ref{eq:sout}) and (\ref{eq:s-tot}).  For the case of
$\lambda=-2/3$, $s_{\rm out}(k)$ grows as $k^{4/3}$, as predicted by
(\ref{eq:sout}), while $s_{\rm in}(k)$ grows slightly faster than
$k^2$, which is consistent with the $k^2\ln k$ growth given in
Eq.~(\ref{eq:sin}).  Notice also that the point where $s_{\rm in}(k)$
and $s_{\rm out}(k)$ intersect moves to larger $k$ as $\lambda$
decreases.  This crossover point also coincides with $k_{\rm max}$ and
the behavior of $k_{\rm max}$ on $\lambda$ is consistent with the
prediction $k_{\rm max}\sim N^{1/(2+\lambda)}$.

\begin{figure}[tbp] 
\includegraphics*[width=8cm]{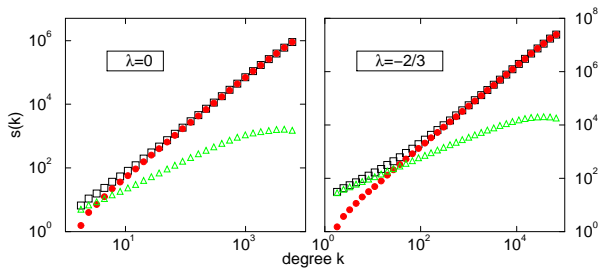}
\caption{The node strength for $\theta=0.5$. The left panel shows the
case $\lambda=0$ and the right $\lambda=-2/3$, both for networks of
size $N=10^6$.  The node strength $s(k)$ and its contribution from
incoming and outgoing links are denoted by $\square,\bullet$ and
$\Delta$ respectively.}
\label{streng_theta}
\end{figure}

While most of our discussion has considered the case where the exponent
$\theta=1$, there are examples of real networks, such as the {\it E coli}
metabolic networks \cite{macdonald} and the airline route network
\cite{orsay}, where the correlations between node degree and link weight seem
to follow $w_{ij}=(k_ik_j)^{1/2}$, that is $\theta = 0.5$.  To test the
predictive power of our analytical approach also for the case of $\theta <
1$, we investigate the scaling of node strength as a function of node degree
for $\theta=1/2$ for the two specific examples of $\lambda=0$ and
$\lambda=-2/3$ (Fig.  \ref{streng_theta}).  Our analytical results for the
outgoing node strength are $s_{\rm out}(k) \sim k$ ($\lambda = 0$) and
$s_{\rm out}(k) \sim k^{5/6}$ ($\lambda = -2/3$), in good agreement with the
best fits of $s_{\rm out}(k) \sim k^{0.9}$ and $s_{\rm out}(k) \sim k^{0.8}$,
respectively.

\begin{figure}[t] 
\vspace*{0.cm}
\includegraphics*[width=8.4cm]{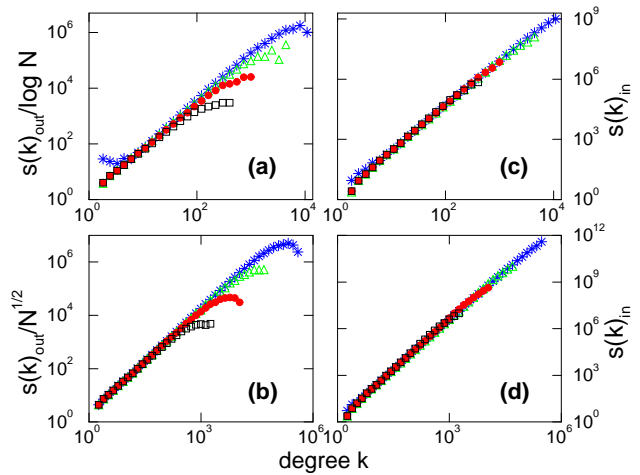}
\caption{Finite size scaling of the node strength contribution
  from outgoing ((a) and (b)) and incoming ((c) and (d)) links for $\lambda =
  0$ (upper panels) and $\lambda = -2/3$ (lower panels).  The network sizes
  are $N=10^4,10^5,10^6$ and $10^7$ ($\square,\bullet,\Delta,\ast$
  respectively).}
\label{strength3}
\end{figure}

As a further test of our theoretical predictions, we show data for the node
strength in which we attempt finite-size scaling (Fig.~\ref{strength3}).  In
panel (a), the data for $\lambda=0$ is consistent with the asymptotic
behavior of $s_{\rm out}(k)/\ln N\sim k^2$ given in Eq.~(\ref{eq:sout}).  The
departure from data collapse arises from the correction term in this
equation.  In panel (b), the data for $\lambda=2/3$ is now consistent with
the form $s_{\rm out}(k)/N^{1/2}\sim k^{4/3}-k^2/N^{1/2}$ given in
Eq.~(\ref{eq:sgen}).  The departure from data collapse is due to the
influence of the correction term in $s_{\rm out}(k)$.  Panel (c) shows the
expected scaling behavior of $s_{\rm in}(k)\sim k^2\ln k$ given in
Eq.~(\ref{eq:sin}).  Curiously, this same behavior also occurs for the case
of $\lambda=-2/3$.

\section{Summary}
 
We studied the statistics of growing networks in which each link $ij$ in the
network has an associated weight $(k_i k_j)^\theta$, where $k_i$ and $k_j$
are the node degrees at the endpoint of the link $ij$.  We also characterized
a node by its strength, defined as the sum of the weights of the links that
are attached to this node.  The link weights and node strengths provide new
metrics with which to characterize heterogeneous networks.  The motivation
for considering these properties stems from a number of real-world examples,
such as airline route networks or metabolic networks, in which the flow
capacity of each link is different.
 
We focused on developing an analytical understanding of the link weights and
node strengths as a function of the underlying growth mechanism of the
network.  Much of our analytical results are obtained by adapting
previously-derived results for node degree correlations to the link weights
and node strengths.  Generically, we found that the total network weight
grows faster than linearly with the total number of nodes $N$.  Strictly
linear growth of the total weight on $N$ occurs for networks whose degree
distribution decays relatively quickly in $k$ for large $k$.  The
distribution of link weights has a power-law tail that is modified by a
logarithmic correction for $\lambda=0$.  There is also a strange and, as of
yet, unexplained anomaly in this distribution when $\lambda<0$.  There are
also small-scale fluctuations in the link weight distributions that are
caused by the granularity in the number of ways that a given link weight can
be factored into a product of the degrees of the endpoint nodes.

A calculational approach in the same spirit as that used for the link weights
was also applied to determine the node strength.  For $\lambda\geq 0$, the
node strength scales as $k^2$, with a logarithmic correction in the case of
$\lambda=0$, while for $\lambda<0$, the node strength has a more complicated
scaling behavior.  The node strength can also be decomposed into
contributions from the outgoing link and a contribution from incoming links.
Only the former is amenable to a complete analytical treatment in terms of
the two-node correlation function, while the latter requires knowledge of
many-node correlations.  We expect that these many-node correlation functions
can be computed withing a rate equation approach.

\section{Acknowledgments}

PLK and SR are grateful to NSF grant DMR0227670 for partial support of this
work.

\appendix
\section{Lower Bound for The Total Link Weight}

Our calculation for the total network weight relied on the distribution
$n_{k,l}$ that was obtained in Ref.~\cite{KR}.  We also made use of the
cutoff $k_{\rm max}$ that applies for the degree distribution $n_k$
\cite{KR-finite}.  Thus the application of this cutoff to $n_{k,l}$ could in
principle be questioned.  Here we establish (for the simplest $\lambda=0$
model) a rigorous lower bound for the total weight of a network using only
the well-known properties of the degree distribution.

Let $W(N)$ be the average total weight of the network with $N$ nodes.  The
next node will link to a target node of degree $k$ with probability
$\frac{1}{2}\,kn_k$.  The degree of the target node then increases $k+1$, so
that the average weight of the new link is $\sum (k+1)\times (kn_k)/2$.  The
increase of the degree of the target node ($k\to k+1$) implies that the
weights of the $k$ current links that end at the target node will also
increase; we denote this corresponding average increase by $(\ldots)$.  Thus
\begin{equation}
\label{WNN}
W(N+1)=W(N)+\sum \frac{1}{2}\,k(k+1)n_k+(\ldots)
\end{equation}
We shall ignore the positive increment $(\ldots)$ since its
calculation requires knowledge of the distribution $n_{k,l}$.  By this
approximation we will derive the lower bound rather than the true
asymptotic.

Equation (\ref{WNN}), in conjunction with the well-known expression
for the degree distribution
\begin{equation}
\label{n_k}
n_k=\frac{4}{k(k+1)(k+2)}\,,
\end{equation}
then gives
\begin{equation}
\label{WNN-log}
W(N+1)>W(N)+\sum \frac{2}{k+2}.
\end{equation}
For the degree distribution, the location of the cutoff $k_{\rm
max}\sim N^{1/2}$ is well established. (Even the full scaling form
$n_k(N)=n_k\,F(k/\sqrt{N})$ is known analytically \cite{KR-finite}.)~
Thus $W(N+1)-W(N)>\ln N$, so that the lower bound for the total weight
is
\begin{equation}
\label{WNN-bound}
W(N)>N\ln N. 
\end{equation}
This bound exhibits a slower growth than the prediction of
Eq.~(\ref{WBA}) but it proves that the total weight of a network grows
faster than linearly in $N$.

For the general case where the attachment rate as the form
$A_k=k+\lambda$, the tail of the degree distribution is given by
$n_k\sim k^{-3-\lambda}$.  For $\lambda>0$ this asymptotic form for
the degree distribution implies that the sum $\sum k(k+1)n_k$
converges, leading to the lower bound $W(N)>N$.  This gives the same
scaling behavior as the actual asymptotic behavior of Eq.~(\ref{WS}).
For $\lambda<0$, the sum $\sum k(k+1)n_k$ diverges and we use the
cutoff $k_{\rm max}\sim N^{1/(2+\lambda)}$ to obtain $\sum
k(k+1)n_k\sim N^{-\lambda/(2+\lambda)}$.  This then gives the lower
bound $W(N)>N^{2/(2+\lambda)}$, a result that also agrees with the
actual asymptotic behavior in Eq.~(\ref{WS}).

\end{document}